\pdfoutput=1
\documentclass[paper,twocolumn]{geophysics}

\usepackage{graphicx}
\usepackage{xcolor}
\usepackage{amsmath,amssymb}

\usepackage{algorithm}
\usepackage{algorithmic}

\usepackage{array}

\usepackage{booktabs}

\usepackage{ftnright}

\makeatletter
\renewcommand{\footnoterule}{%
  \kern -3pt
  \hrule width 0.4\columnwidth
  \kern 2.6pt
}
\makeatother

    \def\bA{{A}}
    \def\bB{{B}}
    \def\Alpha{\boldsymbol\alpha}
    \def\Beta{\boldsymbol\beta}
    \def\bW{{W}}
    \def\bT{{T}}
    \def\bD{{D}}
    \def\bM{{M}}
    \def\bE{{E}}
    \def\bI{{I}}
    \def\bjj{j}
    \def\bee{{e}}
    \def\bQ{{Q}}%
    \def\bbQ{\mathbf{Q}}
    \def\bnull{0}
    \def\calF{\cal F}
    \def\bSigma{\Sigma}
    \def\bGamma{\Gamma}

  \def\loc{{\scalebox{0.6}[0.5]{\rm loc}}}
  \def\cf{{\scalebox{0.7}[0.6]{f}}}
  \def\cn{{\scalebox{0.7}[0.6]{\it n}}}
  \def\cm{{\scalebox{0.7}[0.6]{\it m}}}
  \def\Gauss{{\scalebox{0.6}[0.5]{\rm  Gauss}}}
  \def\GR{{\scalebox{0.6}[0.5]{\rm GR}}}
  \def\Rh{R_{\scalebox{0.7}[0.6]{\rm h}}}
  \def\Rv{R_{\scalebox{0.7}[0.6]{\rm v}}}
  \def\cmpo{{\scalebox{0.7}[0.6]{{\it m}$+1$}}}
  \def\cmmo{{\scalebox{0.7}[0.6]{{\it m}$-1$}}}
  \def\Tr{{\scalebox{0.7}[0.6]{\it T}}}
  \def\src{{\scalebox{0.6}[0.5]{\rm  tx}}}
  \def\rec{{\scalebox{0.6}[0.5]{\rm  rx}}}
  \def\ext{{\scalebox{0.6}[0.5]{\rm  ext}}}
  \def\iso{{\scalebox{0.6}[0.5]{\rm  iso}}}
  \def\subi{{\scalebox{0.6}[0.5]{\rm  sub}}}
  \def\srt{\scalebox{0.7}[0.6]{r;t}}
  \def\str{\scalebox{0.7}[0.6]{t;r}}
  \def\stt{\scalebox{0.7}[0.6]{t;t}}
  \def\sxyz{\scalebox{0.7}[0.6]{x/y/z}}

\newcommand{\tsigma}{\underline{\underline{\sigma}}}
\newcommand{\isigma}{\sigma^\iso}

\makeatletter
\g@addto@macro\normalsize{%
  \setlength\abovedisplayskip{7pt plus 2pt minus 3pt}%
  \setlength\belowdisplayskip{7pt plus 2pt minus 3pt}%
  \setlength\abovedisplayshortskip{2pt plus 2pt}%
  \setlength\belowdisplayshortskip{4pt plus 2pt minus 2pt}%
}
\makeatother
\begin{document}

\title{A Targeted Quadrature Framework for Simulating Large-Scale 3D Anisotropic Electromagnetic Measurements}

\author{
    J{\"o}rn Zimmerling\footnotemark[1],
    Vladimir Druskin\footnotemark[2]\footnotemark[3],
    Sofia Davydycheva\footnotemark[3],
    Wardana Saputra\footnotemark[4],
    Carlos Torres-Verd{\'i}n\footnotemark[4],
    Frank Antonsen\footnotemark[5],
    Jon K\aa re Lotsberg\footnotemark[6],
    and Michael Rabinovich\footnotemark[7]
}

\address{
\footnotemark[1] Department of Information Technology, Regementsvägen 10, Uppsala University, Sweden;
(jorn.zimmerling@it.uu.se).\\
\footnotemark[2] Worcester Polytechnic Institute, USA.\\
\footnotemark[3] 3DEM Holding, USA; (sdavyd@3DEMholding.com).\\
\footnotemark[4] Hildebrand Department of Petroleum and Geosystems Engineering, The University of Texas at Austin, Austin, USA.\\
\footnotemark[5] Equinor, Trondheim, Norway.\\
\footnotemark[6] Equinor, Bergen, Norway.\\
\footnotemark[7] bp, P\&O Subsurface, bp America, Houston, Texas.
}

\footer{Quadrature for 3D EM}
\lefthead{J. Zimmerling, et al.}
\righthead{Quadrature for 3D EM}

\renewcommand{\eqref}[1]{\ref{#1}}

\begin{abstract}

A new efficient and accurate method is presented for simulating frequency-domain borehole electromagnetic (EM) measurements acquired in the presence of three-dimensional (3D) variations of anisotropic subsurface conductivity. The method is based on solving the quasi-static Maxwell equations with a finite-volume discretization via goal-oriented block-quadrature reduced-order modeling. Discretization is performed with a Lebedev grid that enables accurate and conservative solutions in the presence of any form of anisotropic electrical conductivity. Likewise, the method makes use of a new effective-medium approximation to locally account for non-conformal boundaries and large contrasts in electrical conductivity, especially in the vicinity of EM sources and receivers. The finite-volume discretization yields a large symmetric linear system of equations, which is reduced to a set of smaller structured problems via block Lanczos recursion. The formulation also enables the efficient calculation of the adjoint solution, which is necessary for gradient-based inversion of the measurements to estimate the associated spatial distribution of electrical conductivity, i.e., to solve the inverse problem.

The method is applied and verified for borehole ultra-deep azimuthal resistivity (UDAR) measurements commonly used in subsurface geosteering and well navigation. Efficiency, robustness, and scalability are demonstrated using synthetic UDAR data generated in a three-dimensional formation inspired by North Sea geology. The numerical results confirm suitability for integration into real-time UDAR processing frameworks.
\end{abstract}

\section{Introduction}
Electromagnetic (EM) modeling plays a central role in geophysical imaging, where accurate and efficient forward solvers are essential for inversion-based interpretation of remote-sensing measurements. In the frequency domain, and under the quasi-static approximation in the presence of non-magnetic materials  (neglecting displacement currents and assuming that magnetic permeability is a scalar constant $\mu_0$)\footnote[8]{The quasi-static approximation is accurate when displacement currents are small compared to conduction currents, i.e., $\omega\varepsilon \ll \sigma$, where $\varepsilon$ is the permittivity. For the formation resistivities encountered in the applications considered here (up to $100\,\Omega\mathrm{m}$) and frequencies up to $96\,\mathrm{kHz}$, this condition is satisfied with a wide margin.}, Maxwell's equations for the electric field can be recast into a second-order partial differential equation. Specifically, the governing partial differential equation for the electric field  $\mathcal{E}(\mathbf{x},\omega)$ is given by
\begin{equation}
(\nabla \times \,  \mu_0^{-1} \, \nabla \times
+ \imath \, \omega \,  \tsigma(\mathbf{x}) )\, \mathcal{E}(\mathbf{x},\omega)
= - \imath \, \omega \, \mathcal{J}^{\ext}(\mathbf{x}) ,
\label{eq:maxwell}
\end{equation}
where $\tsigma(\mathbf{x})$ is the symmetric positive definite $3\times3$ conductivity tensor, $\omega$ is the angular frequency\footnote[9]{$\exp{(-\imath \omega t)}$ convention in the Fourier transform},  $\mathcal{J}^{\ext}(\mathbf{x})$ denotes an externally applied transmitter current density, and $\imath$ denotes the imaginary unit. The right-hand side of the above equation is fully controlled by the user to interrogate the subsurface. Generally, $\mathbf{x}\in \mathbb{R}^3$ is part of an unbounded domain, and we impose amplitude decay conditions at infinity on $\mathcal{E}(\mathbf{x},\omega)$.

The purpose of geophysical EM imaging is to obtain information about the conductivity tensor $\tsigma(\mathbf{x})$ and its spatial distribution in the subsurface from local measurements of the electric field $\mathcal{E}(\mathbf{x},\omega)$ and/or the magnetic field $-(\imath \omega \mu_0)^{-1} \nabla \times \mathcal{E}(\mathbf{x},\omega)$ at multiple receiver positions $\mathbf{x}_r$, over multiple frequencies $\omega_k$ and for various transmitter--receiver polarizations. We denote such measurements as samples of the form $\mathcal{E}(\mathbf{x}_r,\omega_k)$, collected for a set of user-defined transmitter locations and polarizations.

The associated \emph{forward problem} consists of evaluating the electric field at the receiver locations and frequencies given a conductivity model $\tsigma(\mathbf{x})$ and a set of transmitters. Formally,
\begin{equation}
\big(\tsigma(\mathbf{x}), \mathcal{J}^{\ext}_s\big)
\;\; \mapsto^{\mathbb{F}} \;\;
\Big\{ \, \mathcal{E}(\mathbf{x}_r,\omega_k;\mathcal{J}^{\ext}_s) \Big\}
\label{eq:forward}
\end{equation}
where $\mathbb{F}$ denotes the forward modeling operator, $\{\mathbf{x}_r\}$ are the receiver positions, $\{\omega_k\}$ are the interrogation frequencies, and $\{\mathcal{J}^{\ext}_s\}$ are the enabled transmitters (with position and polarization). The corresponding \emph{inverse problem} seeks to estimate the spatial distribution of the conductivity tensor $\tsigma(\mathbf{x})$ from these measurements.

In this paper, we focus our attention on borehole Ultra-Deep Azimuthal Resistivity (UDAR) measurements, a mo\-dali\-ty widely used in logging-while-drilling (LWD) applications and well geosteering (i.e., well navigation). In typical UDAR surveys, triaxial resistivity tools are employed \citep{Rabinovich2026}, consisting of three orthogonal transmitter polarizations combined with one or more receiver locations that record all three components of the magnetic field. Such multi-polarization configurations are essential to estimate the full conductivity tensor $\tsigma(\mathbf{x})$ of the formation as well as to detect 2D or 3D variations of electrical conductivity. The frequency content used in UDAR applications commonly ranges from approximately $2 \,\mathrm{kHz}$ up to $96 \,\mathrm{kHz}$, enabling depth-of-detection vs. resolution trade-offs that are critical in subsurface well navigation. What differentiates UDAR from other low-frequency borehole measurements is the large separation between transmitters and receivers compared to the bulk skin depth, making it essentially a far-field method \citep{UDarSeydoux}. Although our presentation is tailored to the UDAR case, the numerical simulation method described in this paper is equally applicable to other con\-troll\-ed-source electromagnetic modalities such as CSEM, in particular in the far-field regime (see the discussion of other modalities below).

A forward solver suitable for this class of problems must satisfy several key objectives. It must deliver {\it accuracy}, with high fidelity and tunable tolerances; preserve fundamental {\it physics}, such as reciprocity; and achieve sufficient {\it efficiency} to enable real-time inversion on modern computational accelerators. In addition, it should exhibit strong {\it scalability} for large 3D problems, provide {\it differentiability} through rapid Jacobian computation for gradient-based inversion, and maintain {\it generality} so that a single framework can approach 2D and 3D problems in hierarchical inversion workflows. In this paper, we present a forward modeling algorithm that simultaneously achieves the above objectives in the UDAR setting. We call this algorithm targeted because it approximates the measurements $\mathcal{E}(\mathbf{x}_r,\omega_k;\mathcal{J}^{\ext}_s)$ without computing the field everywhere.

There is a substantial body of literature about algorithms for geophysical EM forward problems and their associated direct and iterative solvers (see, e.g., \citealp{Avdeev2005,haber2014computational,Streich2009,Pardo2021}, and the references therein). In this work, we focus on the so-called reduced-order model (ROM) approaches, which are particularly well suited for large-scale EM problems in geophysics. The central idea of ROM methods in this context is to construct a small-scale (reduced) problem that accurately approximates the EM measurements in large-scale formations---hence the name. Such reduced problems can then be solved efficiently using direct methods, which, due to their poor scalability, are otherwise impractical for large-scale systems. The first method of this class of EM modeling problems was the Spectral Lanczos Decomposition Method (SLDM), introduced by \citet{druskin1987method,druskin1988spectral}. An early account in the Western literature is given by \citet{hoerdt1992thesis}, and a review can be found in \citet{hoerdt1992interpretation}, who analyze the long-offset transient electromagnetic (LOTEM) method, which may be regarded as the surface time-domain analogue of UDAR. This approach solved many practical problems previously considered intractable and enabled reliable me\-dium-scale 3D EM simulations as early as the late 1980s.

In geophysical inversion, the subsurface formation is typically unknown, and the ease and speed of constructing finite-difference (FD) operators often outweigh any accuracy advantages offered by finite-element methods (FEM) \citep[e.g.,][]{Pardo2021}, especially when combined with robust effective-medium approximations based on Kirchhoff’s law. Such an effective-medium construction was advanced in \cite{moskow1999finite} and \cite{druskin1999staggered} to incorporate fully anisotropic conductivities. Moreover, this discretization scheme can be efficiently truncated by introducing optimal exterior coarsening \citep[e.g.,][]{ingerman2000optimal}, effectively simulating an unbounded domain.

While such discretizations are efficient, they also inherit certain challenges intrinsic to diffusive EM modeling. One of the fundamental obstacles in this regime is the null space of the curl-curl operator in equation~\eqref{eq:maxwell}, which consists of curl-free (gradient) fields and leads to spurious modes at low frequencies. Traditionally, this difficulty has been mitigated using vector-potential formulations \citep[e.g.,][]{haber2014computational} with divergence-free gauge conditions, albeit at the cost of increasing both the size and fill-in of the discretized Maxwell system.

The SLDM addresses the above problem by truncating the spurious modes through a spectral correction at a comparatively low computational cost. This, in turn, enables the use of conservative FD schemes on non-uniform Cartesian grids following Lebedev’s formulation in \cite{Lebedev1964}, which is equivalent to a finite-volume (FV) formulation. A better-known special case of such a discretized formulation is Yee's scheme from \cite{yee1966numerical}. We chose the electric-field formulation; owing to the preservation of orthogonality relationships in the discrete operators \citep{Lebedev1964}, it can be equivalently transformed into a magnetic-field formulation (and vice versa) without any loss of accuracy. Thus, the particular choice of formulation is of no consequence.

The SLDM has several limitations. Its convergence can be erratic in the presence of large conductivity contrasts, as shown in \cite{saputra2026udar}. Further, no tight stopping criterion exists for SLDM, it is severely memory-bandwidth-bound on modern computational hardware, and it lacks efficient preconditioners. To address these difficulties, ROM approaches based on rational Krylov subspaces have been proposed as alternatives to SLDM \citep[e.g.,][]{druskin1999new,zimmerling2017model,qiu2019block}, proving competitive for late-time controlled-source EM (CSEM) and other near-zone acquisition methods, though less suitable for far-zone methods.

In this paper, we introduce a new ROM framework that addresses these shortcomings and is tailored to modeling and inversion of UDAR measurements in large-scale formation models. Its two essential features are a two-scale structure and a novel solver based on a goal-oriented block-quadrature approach. The paper makes two main contributions. First, we introduce an SVD-based homogenization that determines the effective direction of conductivity variation in each cell from the data, without prior knowledge of the layering direction. Second, and most importantly, we develop a block-quadrature framework for 3D borehole EM logging \citep[see][]{Zimmerling2025}, which has several advantages over SLDM. It converges in fewer iterations, is more stable, preserves reciprocity exactly, provides a tight computable stopping criterion, and allows a cost-free reduction of the approximation error through quadrature averaging. Because the block formulation works on several vectors at once, it has a higher arithmetic intensity than classical Krylov methods and therefore makes better use of modern hardware, which is limited by memory bandwidth. Because it also yields solutions and adjoint solutions at once, it efficiently computes Jacobians via a two-pass block Lanczos algorithm. Finally, it can be extended with Krein--Nudelman truncators using the equivalent matrix-valued Stieltjes continued fractions representation of block-quadrature \citep{druskin2026}, which can take the formulation beyond classical Krylov convergence.

In the two-scale approach, a finely pixelated formation model is represented on a large-scale (potentially irregular) grid and contains billions of nodes in industrial applications. This model is projected onto a tool-centered Cartesian simulation grid using up to millions of nodes. The simulation grid moves with the tool along realistic 3D curvilinear trajectories. A projection is performed using a novel effective-medium framework, which extracts optimal averaging directions for Kirchhoff's law, corresponding to the principal direction of conductivity-tensor variation within each FV cell computed via principal component analysis, treating the finely pixelized model as the data. This approach is important in pixel-based inversion, when formation boundaries are not sharply defined or are non-layered, unlike in the layered parameterization used in \cite{Davydycheva2003}. Other UDAR inversion codes also use global and local meshes: a global mesh with local refinement along the trajectory stores the conductivity model, and local modeling meshes are generated for each transmitter--frequency pair \citep{Wilson2019}. The two-scale structure can be traced back to the so-called integro-interpolation methods \citep{tikhonov1956difference} developed in the 1950s in the former Soviet Union for nuclear physics simulations. Some elements of such a two-scale model were later implemented in the original SLDM code written at the Central Geophysical Expedition, see~\cite{hoerdt1992interpretation,hoerdt1992thesis}.

The second essential feature for efficient large-scale automated simulation is a robust and scalable Krylov ROM algorithm based on block-quadratures. This approach builds on recent work \citep{Zimmerling2025,druskin2026}, with preliminary applications to UDAR modeling \citep{davydycheva2023modeling,saputra2024adaptive,saputra2026udar}. Block quadrature is a goal-oriented algorithm designed to accurately approximate the multi-input/multi-output (MIMO) transfer function (the EM responses at receiver locations), achieving faster and more stable convergence than SLDM, which computes solutions at all grid points. From the block-quadrature perspective, we derive block-Gauss and block-Gauss-Radau quadratures optimized for far-zone, multi-component electromagnetic arrays.

Compared to SLDM, where termination was often based on ad-hoc self-convergence tests, the difference between the Gauss and Gauss--Radau rules yields a tight error bound that serves as a reliable stopping criterion (see the section ``Comparison with other solvers''), and averaging the two rules reduces the approximation error by an order of magnitude at no computational cost \citep{Zimmerling2025}. In addition, SLDM does not converge monotonically and can behave erratically for hard cases \citep{saputra2026udar}, while the quadrature converges monotonically and is more stable with respect to spurious modes.

Beyond UDAR, the proposed targeted quadrature framework should extend to other modalities with limited observation sets governed by diffusive PDEs, such as geothermal heat flow with sparse temperature arrays or pressure diffusion through fault networks.  This applies in particular to CSEM, where direct solvers are popular because their factorization costs can be amortized over many source positions~\citep{Streich2009}. For large-scale discretizations or sparse survey geometries, targeting specific measurement points and excitation modes allows block quadrature to bypass full-volume solves, which would make it a competitive alternative. Extension to magnetotellurics is possible, but would require truncating the domain at the surface, because Maxwell's equations in the atmosphere are not parabolic. In the UDAR setting, the framework has already enabled fast 2D and 3D hierarchical inversion~\citep{saputra2026rapidOccam} and detailed pre-drill studies~\citep{Rabinovich2026}.

After discussing the components of the proposed quadrature framework in the Methods section, we present three sets of numerical experiments to demonstrate its performance: a 2D test case illustrating the Jacobian approximation, a 3D UDAR log with a curvilinear trajectory using a North Sea model, and a comparison with the open-source library SimPEG.

\section{Method}

This section presents the computational framework for simulating UDAR measurements. We first address the spatial discretization of the governing Maxwell equation~\eqref{eq:maxwell} and the effective medium averaging required for coarse computational grids. Next, we formulate the computation of the UDAR transfer function as a block-quadrature problem. This offers several advantages, most notably the ability to compute both fields and adjoint fields, enabling efficient adjoint-based Jacobian calculations for gradient-based inversion. Finally, we introduce a two-frame coordinate transformation that maps global conductivity models onto local, tool-centered domains to seamlessly handle curvilinear well trajectories.

\subsection{Discretization}\label{sec:Discretization}

\begin{figure}[htbp]
    \centering
    \includegraphics[width=\linewidth]{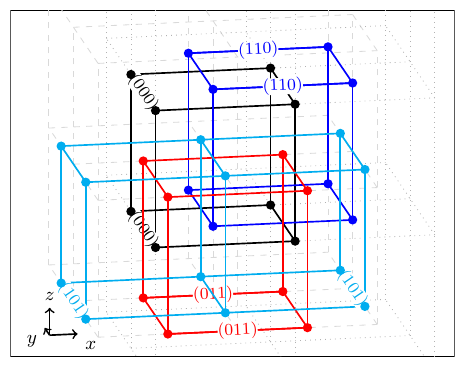}
    \caption{3D view of the four Lebedev clusters shifted in space. Dotted gray lines are the primary grids and dashed gray lines the dual grid.}
    \label{fig:3D_Lebedev}
\end{figure}

Discretizing the quasi-magnetostatic Maxwell equation~\eqref{eq:maxwell} with $N$ degrees of freedom yields the linear system
\[
(\bD + \imath \omega \bM) \widehat{\bee} \;=\; -\imath \omega \, \bjj^{\src}_{\sxyz},
\]
where $\bD$ and $\bM$ are discrete curl-curl and mass/medium operators, respectively, $\bjj^{\src}_{\sxyz}$ is the external forcing current density, and $\widehat{\bee}$ contains the numerical approximation of the electric field.

We use a discretization of the Maxwell operator based on Lebedev's scheme \citep[e.g.,][]{Lebedev1964,jaysaval2016fully}{, although any mimetic discretization that preserves self-adjointness could be used, as evidenced by applying the block quadrature to SimPEG's standard finite-volume discretization in the section ``Comparison with other solvers''}. Standard conservative Yee grids from~\cite{yee1966numerical} are incompatible with anisotropic media because the electric field components are defined on the edges of a Yee cube and are therefore not collocated, which prevents the direct application of tensorial conductivity properties. A Lebedev discretization overcomes this limitation by combining four shifted Yee grids, referred to as clusters, as illustrated in Figure~\ref{fig:3D_Lebedev}. Each cluster is identified by a three-bit binary index $(000), (011), (101)$, and $(110)$, where a ``1'' in a given position indicates a shift along the corresponding axis. Within each cluster, the electric field components are discretized along the edges, as in a conventional Yee grid. By selecting the clusters whose index sum is even, one obtains a consistent Lebedev grid, in which each set of three clusters intersects orthogonally at specific points in space. At these collocation points, all three components of the electric field are defined, allowing the full tensorial medium properties to be applied for accurate modeling.

In summary, each Lebedev cluster supports Yee-grid-style FD differentiation, while the collocation points enable the treatment of fully anisotropic media. The resulting FD scheme remains conservative, preserving key vector identities and physical properties such as current conservation in every elementary cell, as well as symmetry under EM duality transformations (i.e., interchanging the electric and magnetic fields, see \citealp{druskin2007optimal}). The clusters can be grouped into pairs with discretization errors of opposite sign, so their averaging leads to partial error cancellation. To fully exploit this property, the discrete transmitter and receiver current densities (denoted as $\bjj_{\sxyz}^{\src}$ and $\bjj_{\sxyz}^{\rec}$, respectively) are distributed with equal weight over all four Lebedev clusters surrounding the corresponding transmitters and receivers. In UDAR applications, these densities are typically normalized to unit magnetic moments. This discretization enables a reduction of the size of the grid compared to the standard Yee grid, even for isotropic media.

Near the well, we typically employ uniform FD spacing, whereas at domain boundaries, optimal geometric grid coarsening is used \citep[see][]{ingerman2000optimal} to simulate an infinite domain and discretize the exterior problem. The choice of grid spacing and the minimal distance to the geometrically coarsened grid are based on the skin depth relevant to the investigation. Before discretization, we non-dimensionalize and normalize  the Maxwell equations to ensure that the average generalized eigenvalues of the matrix pencil $(\bD,\bM)$ are $\mathcal{O}(1)$ to avoid matrix ill-conditioning (see, e.g., \citealp{Weiss2023}, Sec.~2.2.1) and underflow/overflow in finite-precision arithmetic \citep[see][]{Higham2002}. Using such an FD scheme, $\bM$ is block-diagonal and can be efficiently factorized as $\bM = \bM^{\frac{1}{2}} \bM^{\frac{1}{2}}$ and inverted. The transfer function from transmitters to receivers can be written in terms of the symmetric positive semi-definite matrix
\[
\bA = \bM^{-\frac{1}{2}} \bD \, \bM^{-\frac{1}{2}} \in \mathbb{R}^{N\times N},
\]
namely, the Lebedev approximation of the self-adjoint operator $\tsigma^{-\frac12}\nabla \times \mu_0^{-1} \nabla \times \tsigma^{-\frac12}$.
For example, the transfer function describing the coupling from a $z$-polarized transmitter $\bjj_z^{\src}$ to an $x$-polarized receiver $\bjj_x^{\rec}$ is given by
\begin{equation}\label{eq:couplings}
f^{\srt}_{x;z}(\omega) = -\imath \omega \,
\big(\bM^{-\frac{1}{2}} \bjj_x^{\rec}\big)^{T} \,
(\bA + \imath \omega \bI)^{-1} \,
\bM^{-\frac{1}{2}} \bjj_z^{\src}.
\end{equation}
In UDAR systems with three transmitters and three receivers, nine such couplings need to be computed for every logging point and receiver spacing. In the next subsection, we discuss how to average conductivities to the FD grid to obtain accurate couplings even on coarse grids. We then show that all nine couplings can be efficiently computed using a block formulation.

\subsection{Effective media averaging}\label{sec:EffAveraging}
Using a conservative FD scheme in strongly heterogeneous rock formations requires careful conductivity homogenization in each grid cell, especially when the grid does not align with sharp material interfaces (i.e., non grid-con\-form\-ing media boundaries). The first SLDM code (MAX\-WELL) from~\cite{druskin1988spectral} assumed that conductivity varied locally in a 1D manner along the coordinate axes. This assumption allowed the computation of effective transversely isotropic (TI) FD conductivities using Kirchhoff’s circuit laws, representing the heterogeneous medium as a combination of series and parallel circuits of elementary resistors\footnote{This originated in an unpublished Central Geophysical Expedition (CGE) report by A. Kronrod and A. Lavut, 1979.}. To consider tilted interfaces, \cite{moskow1999finite} proposed layering in arbitrary directions within each cell, producing effective fully anisotropic FD conductivity tensors. However, their homogenization formulas were computationally expensive in cases of internal cross-bedding (such as in eolian or fluvial sandstones, for instance) and, more importantly, required prior knowledge of the effective direction of 1D variation (hereafter referred to simply as the ``effective direction"), which can be difficult to determine in finely pixelated formations derived from EM or other imaging methods.

 Consider the cell $\Omega$ containing (sub-grid) nodes $\mathbf{x}_i \in \Omega$ of the finely pixelated formation grid, with tensor conductivities $\tsigma(\mathbf{x}_i) = \tsigma_i$ for $i=1,\ldots,n^\subi$.

For simplicity, we begin with a scalar (isotropic) conductivity function, interpolated in $\Omega$ as $\isigma(\mathbf{x})$. To determine the effective direction in a data-driven manner, treating the pixelated model as data, we use the representation
\begin{equation}\label{eq:curlcurl}
\nabla\times\nabla\times\mathcal{E}
= -\Delta\mathcal{E} - \nabla\left(\mathbf{c}\cdot\mathcal{E}\right),
\end{equation}
where $\mathbf{c} = \nabla \log (\isigma)$. Here, for simplicity, we omit the explicit dependence on space and frequency. We also use the divergence-free condition for the current density, namely,
\[
\nabla\cdot\left(\isigma\mathcal{E}\right)=0.
\]

The last term on the right-hand side of equation~\eqref{eq:curlcurl}, commonly referred to as the static term, is responsible for spatial discontinuities of $\mathcal{E}(\mathbf{x},\omega)$. We aim to identify a (row) unit vector $\mathbf{k}=(k_x,k_y,k_z)$ with $\|\mathbf{k}\|=1$, representing the effective direction in a finite-volume cell, such that the static term can be approximated via
\[
\isigma(\mathbf{x})\approx\sigma(\mathbf{k}\cdot\mathbf{x}),
\]
where $\sigma(\cdot)$ is a function of one variable, representing conductivity in an equivalent layered medium in the cell. The vector $\mathbf{k}$ is the principal component of $\nabla\log \isigma(\mathbf{x})$ in the cell, representing the dominant direction of variation. This component can be computed as the dominant right singular vector of the singular value decomposition (SVD) of the matrix ${\bGamma}\in \mathbb{R}^{n^\subi\times 3}$ with the rows $\nabla\log \isigma(\mathbf{x}_i)$.

Let $\mathbf{k}_1$ and $\mathbf{k}_2$ form an orthonormal basis in the plane orthogonal to $\mathbf{k}$, so that $\mathbf{k}$, $\mathbf{k}_1$, and $\mathbf{k}_2$ form an orthonormal basis in $\mathbb{R}^3$. Let \[
\hat\sigma=\frac{|\Omega|}{\int_\Omega {[\isigma(\mathbf{x})]}^{-1}d\Omega } \text{ and } \bar\sigma=\frac{\int_\Omega \isigma(\mathbf{x})d\Omega }{|\Omega|} \]
be the harmonic and arithmetic averages, respectively, of the interpolated conductivity in the cell. The harmonic average represents a series circuit (of conductors) along the effective direction, while the arithmetic average represents a parallel circuit along the orthogonal directions. The equivalent finite-\-diff\-er\-ence tensor $\bSigma$ is then defined as
\begin{equation}\label{eq:scalhom}
\bSigma = \mathbf{k}^\Tr \hat\sigma \mathbf{k} + \mathbf{k}_1^\Tr \bar\sigma \mathbf{k}_1 + \mathbf{k}_2^\Tr \bar\sigma \mathbf{k}_2,
\end{equation}
where $\cdot^\Tr$ denotes the transpose.

For anisotropic media, i.e., when $\tsigma(\mathbf{x})$ is a $3\times 3$ symmetric positive-definite matrix-valued function, equation~\eqref{eq:curlcurl} can be approximately\footnote{These formulas are approximations, since generally ${\rm trace}(\tsigma \nabla \mathcal{E}) \neq \tfrac13\,{\rm trace}(\tsigma)\,\nabla \cdot \mathcal{E}$, with equality for isotropic $\tsigma$.} extended to define $\mathbf{k}$, as the principal component of the distribution of $\nabla\,{\rm trace}[\,\log \tsigma(\mathbf{x})]$ in the cell, replacing the scalar $\nabla\log \isigma(\mathbf{x})$ in the isotropic case.

Analogously to equation~\eqref{eq:scalhom}, we obtain the cross-bedding averaging formula
\begin{equation}\label{eq:scalhomcross}
\bSigma = \mathbf{k}^\Tr \mathbf{k}\,\hat\bSigma\, \mathbf{k}^\Tr \mathbf{k}
+ \mathbf{k}_1^\Tr \mathbf{k}_1 \,\bar\bSigma\, \mathbf{k}_1^\Tr \mathbf{k}_1
+ \mathbf{k}_2^\Tr \mathbf{k}_2 \,\bar\bSigma\, \mathbf{k}_2^\Tr \mathbf{k}_2,
\end{equation}
where $\mathbf{k}_1$ and $\mathbf{k}_2$ are as defined above, and $\hat\bSigma$ and $\bar\bSigma$ are the harmonic (series) and arithmetic (parallel) tensorial averages, respectively, of the interpolated  matrix-valued conductivity  $\tsigma(\mathbf{x})$ in $\Omega$.
Except for the SVD  computation of $\mathbf{k}$ in formula~\eqref{eq:scalhom}, this is known in homogenization theory \citep[e.g.,][]{wiener1912theorie,moskow1999finite}. Formula~\eqref{eq:scalhomcross} is algebraically equivalent to the standard homogenization of layered cross-bedded anisotropic media described in the latter paper; however, its matrix expression, more suitable for high-performance computing (HPC) implementations, appears to be new.

\begin{figure}[htbp]
    \centering
    \includegraphics[width=1\linewidth]{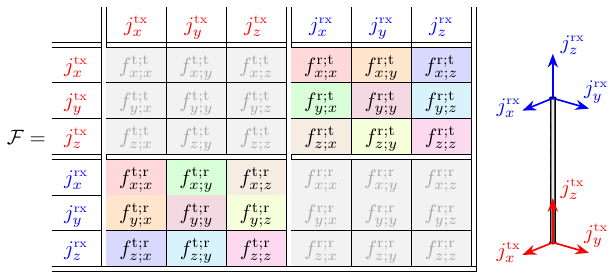}
    \caption{Left: Block-transfer function ${\calF} \in \mathbb{C}^{6 \times 6}$ with 36 couplings, of which 21 are unique due to symmetry (reciprocity). The nine UDAR couplings are in the top-right block and, by reciprocity, the bottom-left block; entries with the same color are equal, and the gray self-couplings are not needed. Right: 3D UDAR measurement system visualizing the three transmitter and receiver current densities.}
    \label{fig:BlockFormualtion}
\end{figure}

\subsection{Block-quadrature formulation}\label{sec:blockQuad}
UDAR's directional triaxial measurements usually require all nine measurement couplings (see equation~\eqref{eq:couplings}) for a given transmitter-receiver separation, i.e., $f^{\srt}_{x;x}, f^{\srt}_{x;y}, \dots $, as shown in the measurement configuration on the right side in Figure~\ref{fig:BlockFormualtion}. At the same time, for gradient-type inversion algorithms, we need to compute Jacobians via an adjoint formulation, which requires computing electric fields on the entire grid for all direct and reciprocal transmitters (i.e., when the receiver dipole acts as a transmitter). The Jacobian calculation does not need to be as accurate as the measurement coupling, since a moderate error in the Jacobian may only slightly slow down convergence of iterative gradient-based inversion without affecting the final result.
Therefore, unlike in the classical SLDM algorithm, we combine all necessary direct and reciprocal solutions for every spacing in one computation, and calculate the UDAR block-transfer function as
\begin{equation}\label{eq:trans}
    {\calF}(\omega)=-\imath \omega \,  \bB^\Tr (\bA+\imath \omega \bI)^{-1}\bB \in \mathbb{C}^{6\times 6},
\end{equation}
i.e., a $6\times 6$ complex symmetric matrix containing all measured nine couplings from transmitters $\bjj^{\src}_{x/y/z}$ to all receivers $\bjj^{\rec}_{x/y/z}$, which are stacked in the block matrix
\begin{equation}
\bB=\bM^{-\frac12}[\bjj^{\src}_x, \bjj^{\src}_y, \bjj^{\src}_z,\bjj^{\rec}_x, \bjj^{\rec}_y, \bjj^{\rec}_z].
\end{equation}
By construction, owing to the symmetry of $\bA$ in Lebedev's discretization and the symmetric block representation in equation~\eqref{eq:trans}, $\calF$ satisfies reciprocity, $f^{\str}_{x;y} = f^{\srt}_{y;x}$, as indicated by the color coding in Figure~\ref{fig:BlockFormualtion}. The figure shows that the relevant couplings can be found in the off-diagonal $3\times3$ block and that the diagonal blocks contain couplings from transmitters and receivers to themselves (e.g., $f^{\stt}$), which are irrelevant. This formulation recasts the problem as an approximation problem for a block-quadratic form as shown in \cite{Zimmerling2025}. This recasting is strictly goal-oriented, i.e., we only require an accurate transfer function (and fields accurate enough for the Jacobians) rather than a good field approximation in the entire domain.

We use a block-Krylov subspace projection of the matrix \( \bA \) onto the space of block-polynomials with degree strictly less than \( m \), i.e.,
\[
\{\, \pi(\bA)\bB : \deg(\pi) < m \,\},
\]
where \( \pi \) is an ordinary polynomial (e.g., \( \pi(z) = c_0 + c_1 z + \cdots + c_{m-1} z^{m-1} \)) evaluated at the matrix~\( \bA \).
The block size \( p \) denotes the number of columns of $\bB$ and is determined by the number of direct and reciprocal transmitters. In UDAR applications, it is \( p = 6 \) if a single spacing is simulated and \( p = 9 \) for two spacings in the same simulation.

Let the columns of ${\bbQ}_\cm=[\bQ_1,\dots,\bQ_\cm]\in \mathbb{R}^{N\times pm}$ contain an orthonormal (${\bbQ}_\cm^\Tr\bbQ_\cm =\bI$) basis for the block Krylov subspace
\[\mathcal{K}_\cm(\bA,\bB)={\rm blkspan} \{\bB, \bA\bB,\dots, \bA^{\cmmo} \bB \}. \]
This subspace can be computed via the block Lanczos iteration, which can be compactly written as
\begin{equation}\label{eq:LancRel}
\bA{\bbQ}_\cm ={\bbQ}_\cm \bT_\cm + \bQ_\cmpo \Beta_\cmpo \bE_\cm^\Tr,
\end{equation}
with  $\bT_\cm$ symmetric positive definite. The block Lanczos iteration is shown in Algorithm~\ref{alg:blockLanc}. Here, we denote by $\bE_\cm \in \mathbb{R}^{mp\times p}$ the canonical block basis vector that is all zeros except for the last $p \times p$ block, which is an identity matrix. In the algorithm, bold Greek letters $\Alpha/\Beta$ represent $p\times p$ matrices, so-called block-coefficients and all capital letters except $\bA,\bT_m$ and $\bE_{1/m}$ are matrices of the size of the block starting vector $\bB$, i.e.~$N \times p$.

The approximation of the block-transfer function from equation~\eqref{eq:trans} on the Krylov subspace reads
\begin{equation}\label{eq:Fm}
  {\calF}\approx {\calF}^\Gauss_\cm =-\imath \omega \Beta_1^\Tr \bE_1^\Tr(\bT_\cm+\imath \omega \bI)^{-1}\bE_1 \Beta_1
\end{equation}
where $\bE_1=[\bI_{p \times p}, {\bnull} ,\dots,\bnull]^\Tr$ is the first canonical block basis vector.

The block coefficients $\Alpha_i$ and $\Beta_i$ are not uniquely determined by the block Lanczos algorithm, because the decomposition of the Gram matrix $\bW^\Tr \bW = \Beta_i^\Tr \Beta_i$ is unique only up to an orthogonal transformation. We choose $\Beta_i$ to be upper triangular, such that the orthogonalization of $\bW=\bQ_i\Beta_i$ can be computed via a thin-QR factorization or QR via Cholesky. We recommend full QR factorization if Jacobians need to be computed and QR via Cholesky to reduce computational time if only responses $\calF$ are needed. With this choice, $\bT_\cm$ is a banded matrix with bandwidth $p$ and ${\calF}^\Gauss_\cm$ can be efficiently computed using complex banded solvers (e.g., \texttt{ZGBSV}; \citealp{lapack_users_guide}).

For certain operators associated with uniform measures, the non-block version of the algorithm coincides with the classical Gaussian quadrature used to approximate integrals. The block extensions of this formulation via the block Lanczos algorithm then lead to block-\-Gauss-\-qua\-dra\-ture \citep[see, e.g.,][]{golub_meurant_2010}. Adding $p$ quadrature points at zero by modifying the last block diagonal element of $\bT_\cm$ to ${\hat\Alpha_\cm }=\Alpha_\cm+\Delta \Alpha_\cm $ leads to a block-Gauss-Radau quadrature approximation $\widehat{\calF}_\cm^{\GR}$ given by
\begin{equation*}
\widehat{\calF}_\cm^{\GR} =- \imath \omega \Beta_1^\Tr \bE_1^\Tr(\bT_\cm+\bE_\cm \Delta \Alpha_\cm \bE_\cm^\Tr+\imath \omega \bI)^{-1}\bE_1 \Beta_1.
\end{equation*}
In \cite{Zimmerling2025}, it was shown that the difference between the Gauss and Gauss-Radau quadratures $\|{\calF}_\cm^{\Gauss}-\widehat{\calF}^{\GR}_\cm\|$ is a tight error bound for the approximation error $\|{\calF}-{\calF}^{\Gauss}_\cm\|$, which we use as a stopping criterion of the block Lanczos iterations. Further, it was shown that averaged quadrature rules ${\scriptstyle \frac{1}{2}}[{\calF}_\cm^\Gauss+\widehat{\calF}_\cm^\GR]$ lead to an order-of-magnitude reduction in the approximation error. Additional computational advances incorporating the continuous nature of the spectrum of the Maxwell operator in unbounded domains are discussed in \cite{druskin2026}.

Unlike SLDM, the block-Gauss and block-Gauss-Radau approximations preserve EM reciprocity by construction, i.e., ${\calF}_\cm$ is complex symmetric and thus invariant under swapping transmitter and receiver.

 \begin{algorithm}[htbp]
\caption{Block Lanczos iteration}\label{alg:blockLanc}
\begin{algorithmic}[1]
\STATE Given $m$, and $\bA\in{\mathbb R}^{N\times N}$
, $\bB\in{\mathbb R}^{N\times p}$
\STATE Orthonormalize $\bB$,  as $\bQ_1 \Beta_{1} = \bB$
\STATE $\bW \gets \bA\bQ_1$
\STATE $\Alpha_1 \gets \bQ_1^\Tr \bW$
\STATE $\bW \gets \bW - \bQ_1 \Alpha_1$
\FOR{$i= 2,\dots, m$}
 	\STATE Orthonormalize $\bW$, such that $\bQ_i\Beta_{i}=\bW$
	\STATE $\bW \gets \bA\bQ_i- \bQ_{i-1}\Beta_{i}^\Tr$
 	\STATE $\Alpha_i \gets \bQ_i^\Tr W$
 	\STATE $\bW \gets \bW - \bQ_i \Alpha_i$
\ENDFOR
\STATE Form $\bT = \begin{bmatrix}
\Alpha_1 & \Beta_2^\Tr & & \\
\Beta_2 & \Alpha_2 & \ddots & \\
& \ddots & \ddots & \Beta_{\cm}^\Tr \\
& & \Beta_{\cm} & \Alpha_\cm
\end{bmatrix}$
\end{algorithmic}
\end{algorithm}

The block-quadrature algorithm generalizes to other matrix functions, i.e., if block time-domain signals are required, they can be efficiently approximated as
\begin{equation*}
\bB^\Tr \exp{(-\bA t)}\bB \approx \Beta_1^\Tr \bE_1^\Tr\exp{(- \bT_\cm t)} \bE_1 \Beta_1.
\end{equation*}
We stress that the block-quadrature algorithm is an all-at-once algorithm, i.e., it gives a solution for all frequencies $\omega$ or all times $t$ with a single run of the algorithm.

Due to round-off errors, spurious curl-free modes may appear in numerical computations. These modes manifest as small, non-physical eigenvalues of $\bT$, which must be removed in the SLDM framework via spectral truncation  \citep[e.g.,][]{druskin1994spectral}. While this procedure is inexpensive for serial computations, it can become a bottleneck on modern CPUs because of limited parallelization. In contrast, within the quadrature-based approach, even this step becomes unnecessary, since the associated spurious spectral weights, i.e., the residues of $\Beta_1^\Tr \bE_1^\Tr (\bT_\cm + \imath \omega \bI)^{-1} \bE_1 \Beta_1$ as a function of $ \imath \omega$, are on the order of machine precision and therefore have a negligible effect.

A drawback of Krylov model reduction, including qua\-dra\-ture-based approaches, is that
classical preconditioners cannot be used. However, randomized Ny\-ström-pre\-con\-di\-tion\-er approaches show potential \citep[e.g.,][]{chen2025Nystrom,Zimmerling2025}, especially for GPU implementations, and this remains an area of future research.

The block recursion offers several computational advantages over both direct solvers and standard (non-block) Krylov methods.
Compared to direct solvers, the block recursion is very memory efficient, since only three block vectors need to be stored in memory (about \texttt{144 MB} per million unknowns). Further, direct solvers have an asymptotic complexity of $\mathcal{O}(N^2)$ for 3D problems and remain computationally intractable for $N$ in the millions of unknowns.

Using FD discretizations, the matrix--\-block-\-vector product $\bA \bQ_i$ can be implemented in a matrix-free fashion; on modern hardware, this step is memory-bandwidth limited and dominated by loading the averaged conductivities and block vectors.  At the same time, working with block vectors rather than individual vectors, as in ordinary Krylov methods such as SLDM, increases the arithmetic intensity of the algorithm and thus makes better use of current computing systems, which are memory-bandwidth-bound for these algorithms. When $p$ is small, the computational cost is dominated by matrix--vector products $\bA \bQ_i$ and increases affinely in $p$. The block method reduces the number of iterations required to reach a given tolerance by about a factor of two compared to SLDM, and together with averaged quadrature rules and a tight stopping criterion, it avoids unnecessary iterations and accelerates the computations. In the future we plan to accelerate the quadrature solver using a GPU implementation, randomized preconditioners, and Krein--\-Nudel\-man approaches to block quadratures \citep[see][]{druskin2026,karpelin2025fast}.

In summary, compared to SLDM, the block-quadrature formulation is numerically more stable, better suited to modern hardware, and converges in fewer iterations; it preserves reciprocity exactly and, as shown in the next subsection, is well suited for adjoint-based computation of Jacobians.

\subsection{Efficient computation of Jacobians in the self-adjoint formulation via block Lanczos}\label{sec:Adjoint}
For gradient-based inversion, we need to compute the sensitivity $\delta {\calF}$ of the measurements $ {\calF}$ to perturbations in the conductivity tensor $\delta \tsigma$. Perturbations in the conductivity model after discretization become perturbations in the medium matrix $\delta \bM$. We use the adjoint-state method \citep[e.g.,][]{ReneAdjoint} to compute elements of the Jacobian, which requires the electric field originating from the transmitters, and the adjoint field caused by the receivers; however, we use the self-adjoint formulation of the Maxwell operator, eliminating the need to solve two separate equations for fields and adjoint fields. The elements of the Jacobian are the sensitivities of the measurements $\delta {\calF}(\omega)$ with respect to perturbations of the material coefficients in the mass/medium matrix $\delta \bM$. A straightforward derivation yields the directional (Fréchet) derivative
\begin{equation*}
\begin{aligned}
  \delta {\calF}(\omega) &= - \omega^2 \, [(\bA+\imath \omega \bI)^{-1}\bB]^\Tr \, \bM^{-\frac12} \, (\delta\bM)  \cdot \\
                    &\quad \bM^{-\frac12} (\bA+\imath \omega \bI)^{-1}\bB.
\end{aligned}
\end{equation*}
Because we choose a self-adjoint formulation ($\bA=\bA^\Tr$) of our problem, we observe that $\widehat \bE=-\imath \omega\bM^{-\frac12}(\bA+\imath \omega \bI)^{-1}\bB$ are electric fields\footnote{The $\widehat \cdot$ is used to visually distinguish the electric field approximation $\widehat{\bE}_\cm^{\Gauss}$ from $\bE_\cm$ a block-canonical basis vector.} and adjoint fields at the same time, leading to succinct Jacobian computation via
\begin{equation}
    \delta {\calF}(\omega)= [\widehat \bE(\omega)]^\Tr  \, \delta \bM\, \widehat \bE(\omega).
\end{equation}
Computation of these inner products is fast for perturbations $\delta \bM$ of small support. The block-quadrature formulation is naturally suited to adjoint methods, since $\bB$ contains the transmitters and receivers. Thus, $\widehat \bE(\omega)$ can be approximated in the (scaled) block-Krylov subspace for all transmitters and receivers as
\[
\widehat{\bE}(\omega) \approx\widehat \bE_\cm^\Gauss(\omega) = -\imath \omega \bM^{-\frac12} \bbQ_\cm (\bT_\cm+\imath \omega \bI)^{-1}\bE_1 \Beta_1.
\]
The above (Gauss) field and adjoint field approximations can again be improved by averaging with Gauss-Radau approximations. Storage of the block Lanczos vectors $\bbQ_\cm $ is infeasible and would negate the memory advantage of these methods; however, $\widehat \bE^\Gauss_\cm(\omega)$ can be computed by running the block Lanczos algorithm twice to expand $\widehat \bE^\Gauss_\cm(\omega)$ in the second run from the known block-expansion coefficients $(\bT_\cm+\imath \omega \bI)^{-1}\bE_1 \Beta_1$, so-called two-pass Lanczos. The computational cost of the Jacobian computation is thus roughly twice the cost of approximating the responses ${\calF}_\cm$.

\subsection{Two-frame coordinate model for well-path and formation mapping}\label{sec:Curvelinear}
The formation at each point in the 3D space is specified in a global (Earth-related) coordinate system (x, y, z) by the vertical and horizontal resistivities, $\Rv$ and $\Rh$, respectively, assuming TI-anisotropy.   We allow an arbitrary tilt of the formation anisotropy tensor, with formation dip, $\theta_\cf$, and azimuth, $\phi_\cf$, by defining the local stratigraphic vertical direction of anisotropy, which may vary spatially. Thus, $\Rv$ and $\Rh$ need not be strictly aligned with the global vertical and horizontal directions. This representation, consistent with a locally layered medium, is simpler than the most general anisotropic case and avoids the need for full-tensor descriptions in coordinate-transformation equations. Specifically, we parameterize the conductivity at each point with four unknowns, whereas a general symmetric tensor would require six.

The well path through the formation follows a curvilinear trajectory, with position $(x_\cn, y_\cn, z_\cn)$ in a right-handed coordinate system (with the $z$-axis directed downward, e.g., North-East-Down), and orientation defined by the dip $\theta_\cn$ and azimuth $\phi_\cn$ at each logging point $n$. This path is usually computed using the minimum-curvature method, although such an approach is not strictly necessary.%

Our computational grid is aligned with and centered on the tool, such that a coordinate transformation is performed at each logging point from the global coordinates ($x, y, z$) to the local coordinates ($x_\cn^\loc, y_\cn^\loc, z_\cn^\loc$) related to the tool. The Lebedev grid, attached to the transmitter and receiver(s), remains identical at each point and moves with the tool along the trajectory through the formation.

An example of a global model with a curvilinear trajectory, together with the corresponding tool-centered grid, is shown in the Measurement example.

The local coordinate frame at each logging point $n$ is tangential to the well path (i.e., aligned with the tool). First, the coordinate system is translated to $(x_\cn, y_\cn, z_\cn)$, then rotated by the azimuth angle $\phi_\cn$ about the $z$-axis (in the $xy$-plane), and finally by the dip angle $\theta_\cn$ about the new $y$-axis (in the $xz$-plane), such that the local $z$-axis is aligned with the tool. Both rotations are right-handed about the positive axes; since $z$ points downward, the azimuth is measured clockwise from north when viewed from above.

It is within this local coordinate frame that the grid is constructed, and the formation properties are sampled and averaged. The formation dip and azimuth, meanwhile, are transformed (at each point in 3D space) to a local formation dip and azimuth by compounding the global formation parameters and local coordinate rotations. These are standard Euler angle rotations, which are common in anisotropic EM modeling \citep[e.g.,][]{jaysaval2016fully,hansen2016tti}; we state them here for completeness.

With $\theta_\cf$ and $\phi_\cf$ denoting the formation dip and dip azimuth, the downward bedding normal is $(-\sin\theta_\cf\cos\phi_\cf,\,$ $-\sin\theta_\cf\sin\phi_\cf,\, \cos\theta_\cf)$. Expressed in the local frame, the bedding orientation is given by
\begin{center}
{
\noindent
{$\begin{aligned}
\theta_{\cf,\cn}^\loc &= \arccos\!\Bigl(
\cos\theta_\cf\cos\theta_\cn
- \sin\theta_\cf\cos(\phi_\cf-\phi_\cn)\sin\theta_\cn
\Bigr),\\
\phi_{\cf,\cn}^\loc &= \operatorname{atan2}\!\Bigl(
\sin\theta_\cf\,\sin(\phi_\cf-\phi_\cn),\\
&\qquad \sin\theta_\cf\,\cos(\phi_\cf-\phi_\cn)\,\cos\theta_\cn + \cos\theta_\cf\,\sin\theta_\cn
\Bigr),
\end{aligned}
$}}
\end{center}
where $\theta_{\cf,\cn}^\loc$ is the angle between the tool axis and the downward bedding normal, $\phi_{\cf,\cn}^\loc$ is the azimuth of the upward bedding normal around the tool axis, measured from the high side of the tool toward its right side, and $\operatorname{atan2}$ denotes the four-quadrant arctangent. In each cell of the Lebedev grid, a material averaging of the formation properties is performed, and the averaged conductivity tensor is computed.

\subsection*{Overview of Methods}%
The main technical contribution of our work is a forward modeling method for EM geophysical surveys, verified in the UDAR setting, that meets the objectives outlined in the introduction. Our approach combines a {\it scalable} discretization scheme with {\it accurate} effective-medium averaging to minimize grid dimensions. A new, {\it efficient} and {\it scalable} block-quadrature solver, tailored to UDAR and modern computational hardware, delivers significant speedups while preserving key {\it physics} such as reciprocity. It also offers {\it differentiability} via an adjoint-based Jacobian framework, enabling gradient-based inversion. Finally, the design of both algorithm and discretization is sufficiently {\it general} to accommodate 2D and 3D problems within hierarchical inversion workflows.

\begin{figure}[htbp]
    \centering
    \includegraphics[height=0.194\textwidth]{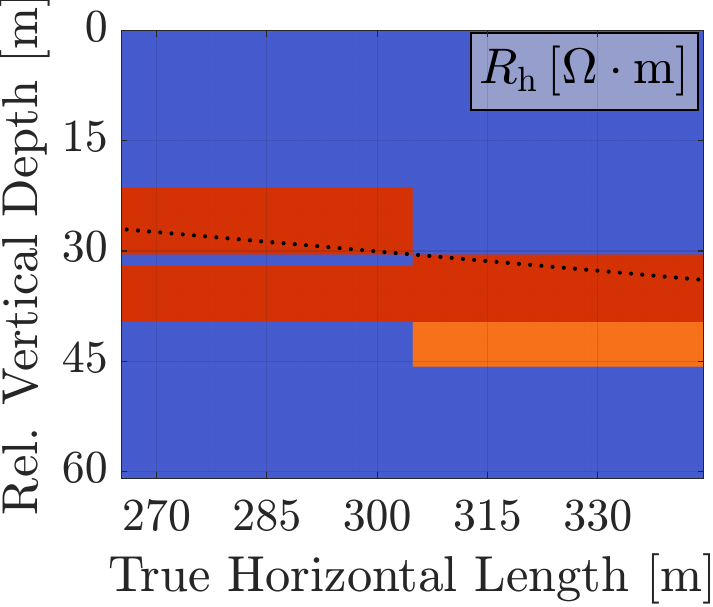}
    \includegraphics[height=0.190\textwidth]{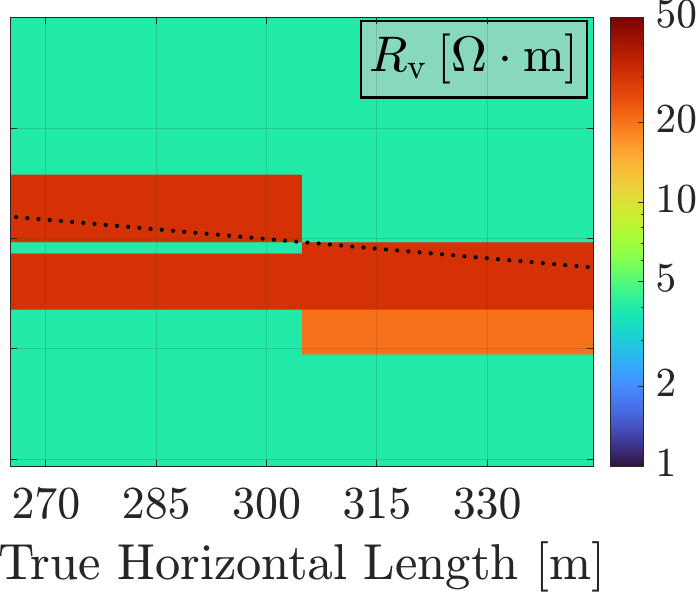}
    \caption{Slice of a layered resistivity model with a fault. The horizontal resistivity, $\Rh$ in $[\Omega \cdot m]$, is shown on the left and the vertical resistivity, $\Rv$, on the right.}
    \label{fig:true_Rh}
\end{figure}

\begin{figure}[htbp]
    \centering
    \includegraphics[width=\columnwidth]{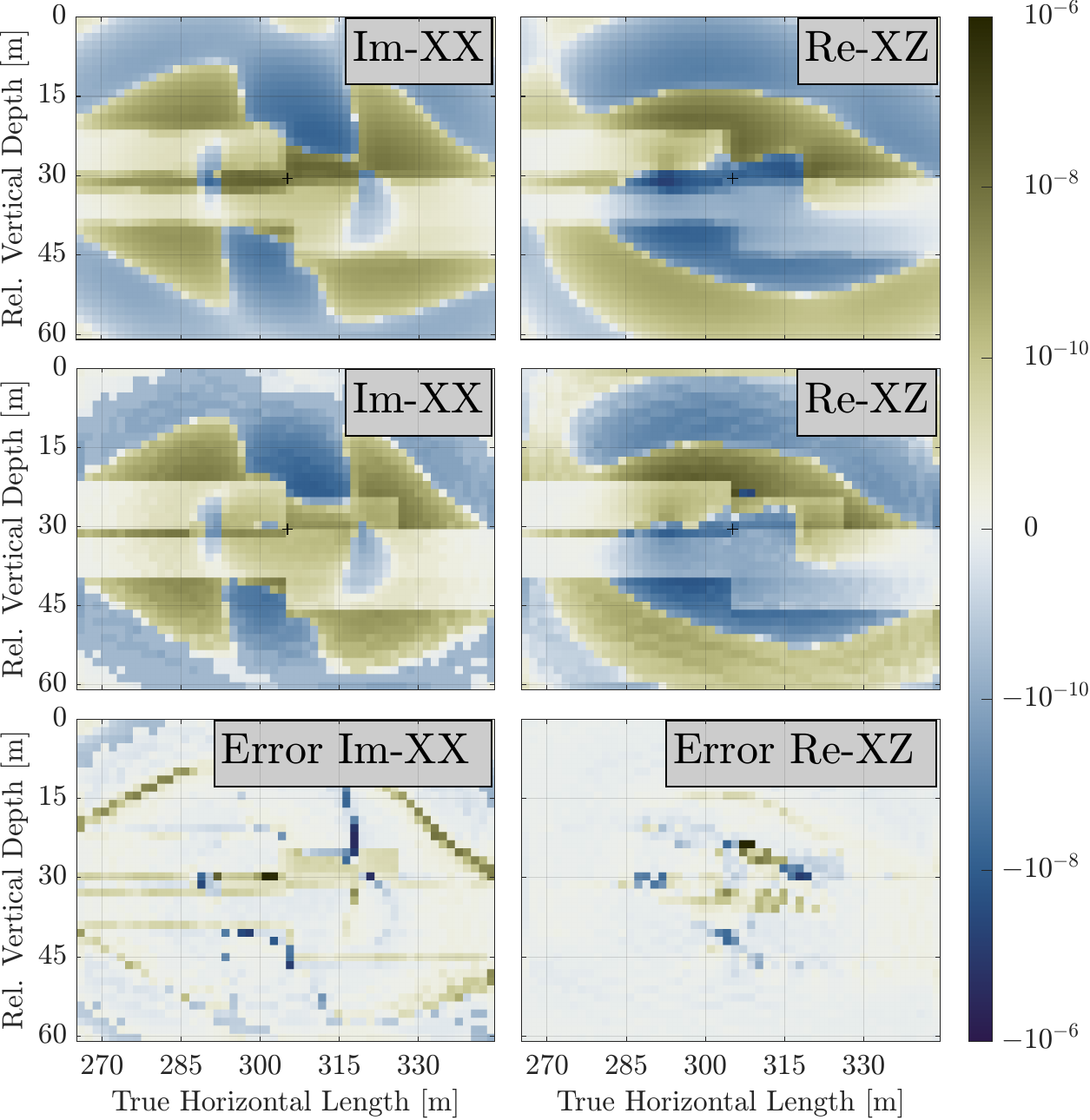}
    \caption{Example components of the adjoint Jacobian computation for $\Rh$ (top-row) compared to their FD counterparts (middle-row) and their error (bottom-row).}
    \label{fig:jac1}
\end{figure}

\section{Results}
We present three numerical examples. First, a small-scale 2D example demonstrates the Jacobian computation. Second, a 3D synthetic UDAR experiment in a complex formation is simulated as a self-consistency test, followed by a discussion of the HPC implementation and computational performance. Third, the next section compares the proposed solver with the open-source library SimPEG in terms of speed, memory, and accuracy on a separate isotropic benchmark model.

\subsection{Jacobian computation}\label{sec:JacExample}
We use a small anisotropic fault model to verify the computation of the Jacobian using the presented solver. Figure~\ref{fig:true_Rh} (left) shows the horizontal resistivity distribution, while the right panel shows the vertical resistivity. The configuration is a small section of a larger model starting at a true vertical depth (TVD) of 5.2~$k$m and a true horizontal length (THL) of 265~m. In all figures, we use relative vertical depth starting from a TVD of 5.2~$k$m. The well (dotted line) intersects a vertical fault, with $90^{\circ}$ dip and zero azimuth. Although the model is parameterized as a diagonally anisotropic medium, material averaging in a tool-centered coordinate system results in full tensorial resistivities~\citep[see][]{5991949,Lee2010}. For the synthetic model, we simulate a tool with two receiver spacings of $25.3$ and $13.1\,\mathrm{m}$, with operating frequencies of $(6,\, 12,\, 24)\,\mathrm{kHz}$ and $(24,\, 48,\, 96)\,\mathrm{kHz}$, respectively.

For this simple anisotropic global model consisting of ($52 \times 1 \times 40$) pixels, with 52 logging points and 6 tool combinations, 1,297,921 forward simulations ($n_m+1$) are required to calculate the Jacobian using FD differentiation. On a typical desktop computer without parallelization, such a computation would take more than 50~days to complete, making brute-force sensitivity computation infeasible for 2D or 3D inversions. In contrast, using the adjoint method, the sensitivity with respect to $\Rh$ can be computed in approximately 20~seconds (using 104 MPI tasks and 4 OpenMP threads per rank), representing a substantial reduction in computational cost.

\begin{figure}[htbp]
    \centering
    \includegraphics[width=\columnwidth]{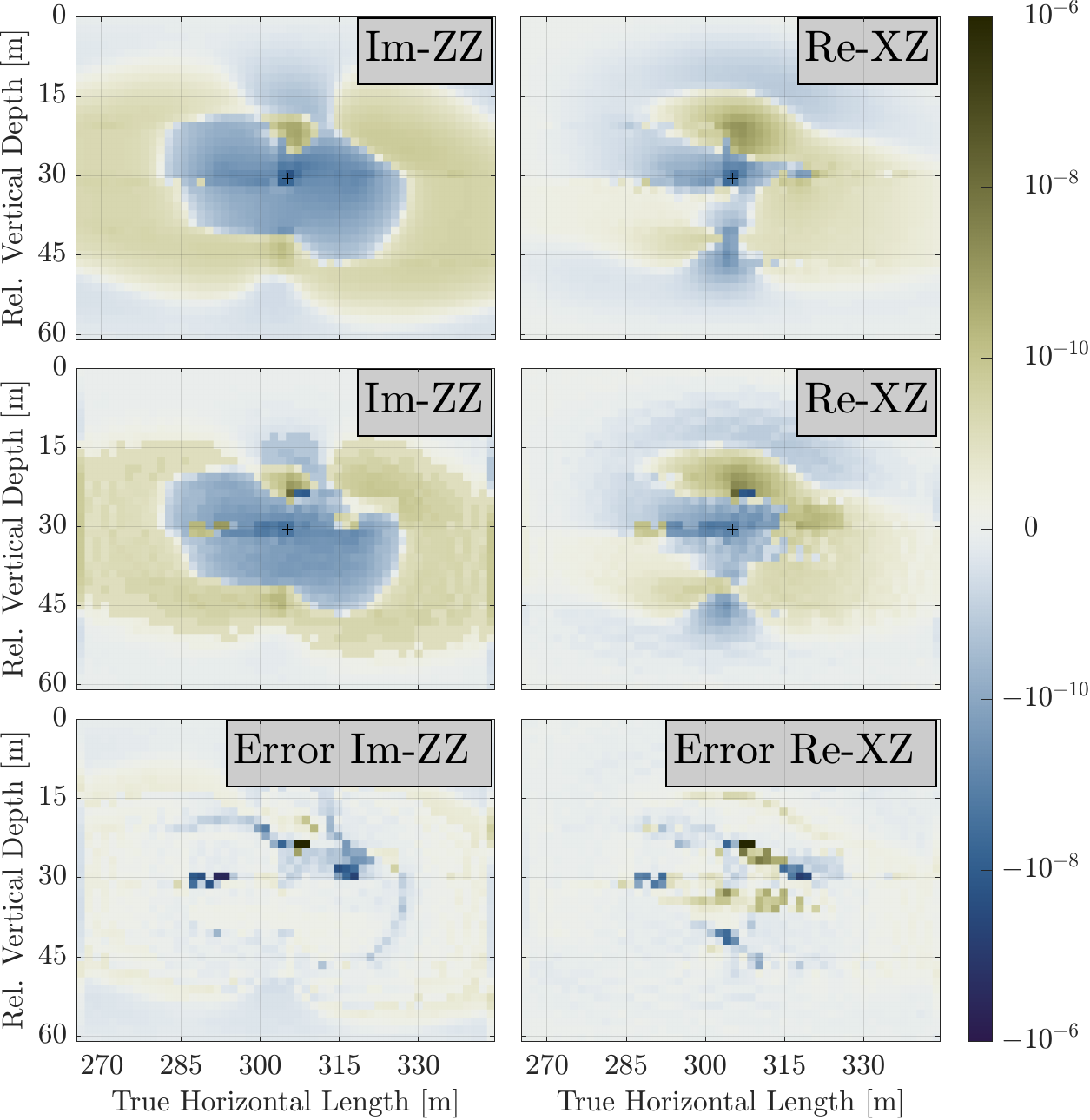}
    \caption{Example components of the adjoint Jacobian computation for $\Rv$ (top-row) compared to their FD counterparts (middle-row) and their error (bottom-row).}
    \label{fig:jac2}
\end{figure}

 Figures~\ref{fig:jac1} and~\ref{fig:jac2} compare the sensitivities with respect to perturbations in $\Rh$ and $\Rv$, respectively. The top rows show the adjoint-based sensitivities, the middle rows show the corresponding elements of the Jacobian computed via finite-difference differentiation, used to verify our adjoint-based Jacobian formulation, and the bottom rows show the error between the two. While the method computes sensitivities for all nine couplings and all frequencies simultaneously, for brevity we display here only selected real and imaginary components.

 The sensitivities agree well with the finite-\-diff\-er\-ence results, which show slight numerical noise. The noise arises from the fact that the transfer function is only locally linear with respect to small perturbations in the medium. For points far from the transmitter and receiver location, this requires numerical tolerances near double precision round-off for accurate finite-\-diff\-er\-ence evaluation. Because the Jacobian entries vary over several orders of magnitude and change sign, relative errors are not meaningful near zero crossings. We therefore display the error of the adjoint and finite-difference Jacobians. The errors are largest near the roots of the Jacobian and at structural interfaces.

\begin{figure*}[htbp]
    \centering
    \includegraphics[width=\linewidth]{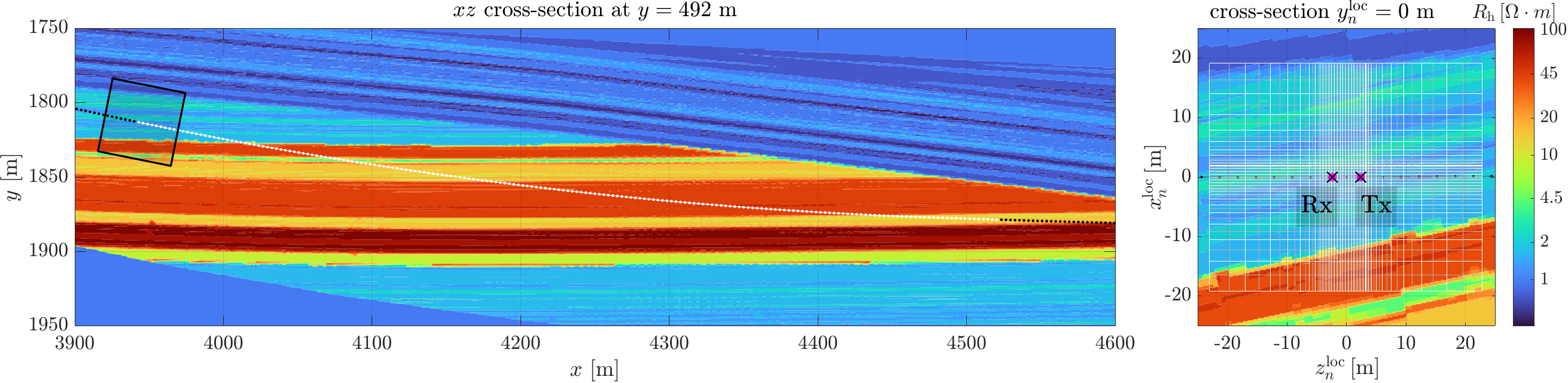}
    \caption{Left: Vertical cross-section of a full 3D synthetic reservoir model with general anisotropy intersected by a curvilinear well trajectory (black and white dotted lines, respectively, behind and in front of the model cross-section). Right: FD grid attached to the tool’s transmitter and receiver (black $\boldsymbol{\times}$), on the background of the 3D medium cross-section (see black shaded rectangle on the left for location). White and gray lines identify primary and dual nodes, respectively, of the Lebedev grid.}
    \label{fig:GlobalModel}
\end{figure*}

\newcommand{\scconst}{0.35}

\begin{figure*}[htbp]
    \centering
    \includegraphics[height=\scconst\textwidth]{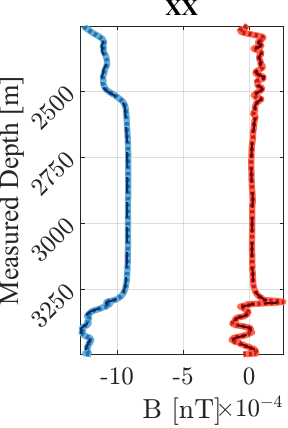}
    \includegraphics[height=\scconst\textwidth]{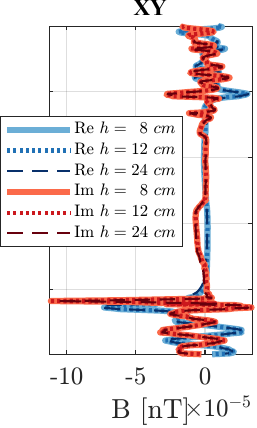}
    \includegraphics[height=\scconst\textwidth]{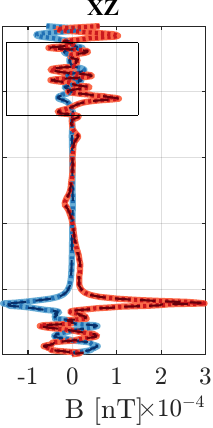}
    \includegraphics[height=\scconst\textwidth]{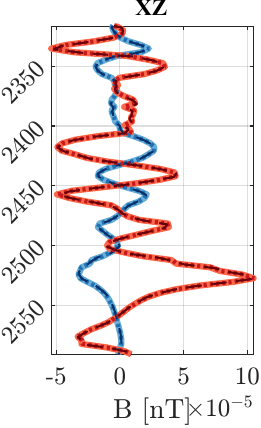}
    \caption{Synthetic UDAR measurements at $12~\mathrm{kHz}$ through the medium described in Figure~\ref{fig:GlobalModel}. Displayed are three couplings XX, XY and XZ of the B-fields (real, in blue, and imaginary parts, in red). Solid, dotted and dashed lines correspond to the uniform grid steps of, respectively, 8, 12, and $24\, \mathrm{cm}$. The box in the XZ graph indicates the landing section that is shown with more detail in the rightmost graph.}
    \label{fig:all12kHz}
\end{figure*}

\begin{figure*}[htbp]
    \centering
    \includegraphics[height=\scconst\textwidth]{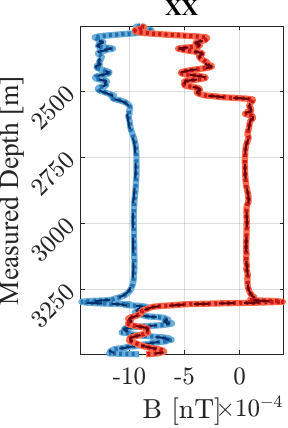}
    \includegraphics[height=\scconst\textwidth]{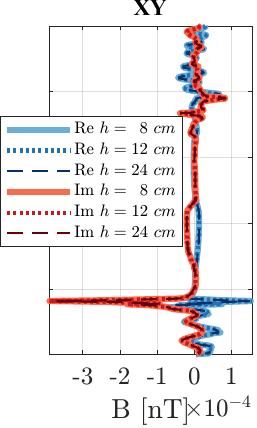}
    \includegraphics[height=\scconst\textwidth]{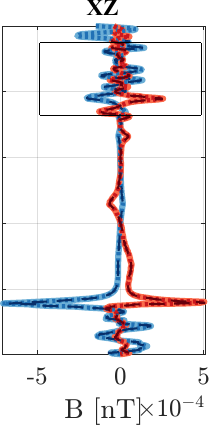}
    \includegraphics[height=\scconst\textwidth]{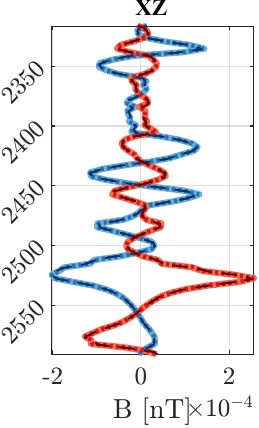}
    \caption{Synthetic UDAR measurements at $48~\mathrm{kHz}$ through the medium described in Figure~\ref{fig:GlobalModel}. The landing section, as indicated by the black box, is shown with more detail in the rightmost graph.}
    \label{fig:all48kHz}
\end{figure*}

\subsection{Measurement example}\label{sec:3DLogs}
The developed simulation method was tested and verified by comparison with independent 1D and 3D modeling results \citep[see][]{saputra2024adaptive,saputra2026udar}{, and by the comparison with SimPEG and a 1D semi-analytical solution in the section ``Comparison with other solvers''}.
Here, we describe a forward-modeling self-consistency test using a challenging 3D synthetic model inspired by a realistic borehole trajectory, transmitter-receiver configurations, and an anisotropic electrical conductivity as encountered in the North Sea.

Figure~\ref{fig:GlobalModel} on the left shows a vertical cross-section of the synthetic reservoir model intersected by a curvilinear well trajectory, indicated with a white dotted line in front of the section and a black dotted line behind it. The model represents a rotated and eroded fault block characteristic of North Sea hydrocarbon traps. Reservoir zones, shown in red and yellow, correspond to hydrocarbon-filled formations. The simulated well trajectory is designed to assess the sensitivity of UDAR measurements to fault proximity, evaluating how accurately faults can be mapped for geosteering decisions.

There are a total of 408 logging points along the well trajectory. Figure~\ref{fig:GlobalModel} on the right shows the FD grid attached to the tool's transmitter and receiver, overlaid on the medium cross-section. The grid is generated automatically, with a uniform discretization used between the transmitter and receiver and around the tool. The uniform region is centered on the tool and extends over about 1.5 transmitter--receiver spacings along the tool axis and about one spacing across it. The outer regions are terminated by 10--20 geometrically increasing, spectrally optimal steps (growth factor of about 1.4, with more steps for finer meshes) to emulate an infinite domain, such that the half-width of the full grid, including the uniform region, is four transmitter--receiver spacings. Typically, between 10 and 20 uniform nodes per minimal skin depth are required for a given model and tool frequency range; the coarsest discretization deliberately falls below this.

A single run of the forward-modeling algorithm computes all nine magnetic couplings at all tool frequencies for one logging position.
Figures~\ref{fig:all12kHz} and~\ref{fig:all48kHz} show the synthetic logs at $12~\mathrm{kHz}$ and $48~\mathrm{kHz}$, respectively. For brevity, only the XX, XY, and XZ couplings are displayed. The left three graphs show the entire set of measurements, while the right-most graph is a zoomed-in view of the landing section between 2300 and 2600\,m measured depth (MD, the length along the well path), indicated by the black box in the XZ panel. Real and imaginary parts are shown in blue and red, respectively. Solid, dotted, and dashed lines correspond to uniform grid spacings of 8, 12, and $24\,\mathrm{cm}$, respectively (60, 40, and 20 uniform steps per transmitter--receiver spacing).

Minimal and maximal skin depths for the given medium model and frequency range are $1.4\,\mathrm{m}$ and $54\,\mathrm{m}$, respectively. The two finest discretizations agree closely, with only minor discrepancies observed for the coarsest grid. This example verifies that the developed effective-medium formulation accurately captures the EM response even at relatively coarse discretizations.

\begin{figure}[htbp]
    \centering
    \includegraphics[width=0.9\columnwidth]{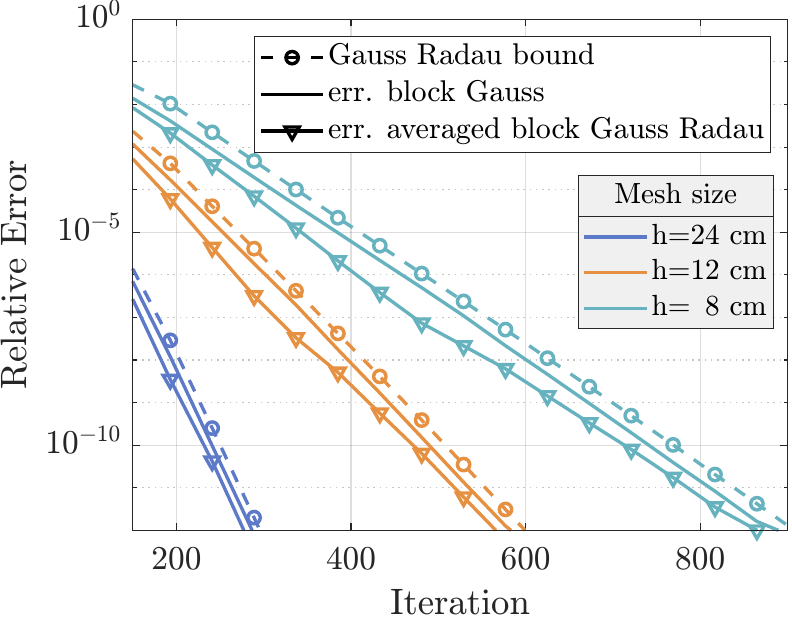}
    \caption{Convergence of the solver for a single logging point in the landing zone for the three different mesh sizes.}
    \label{fig:Convergence}
\end{figure}

The three discretizations result in systems of increasing sizes $N_{24\,\mathrm{cm}} = 320\times10^3$, $N_{12\,\mathrm{cm}} = 516\times10^3$, and $N_{8\,\mathrm{cm}} = 1058\times10^3$ unknowns, respectively. Figure~\ref{fig:Convergence} shows the solver convergence for a logging point located within the landing zone. Displayed is the relative error in the UDAR responses corresponding to the off-diagonal $3\times3$ block of ${\calF}$. The number of iterations required to reach a given tolerance is approximately inversely proportional to the minimum mesh size,~$h$.

The latter behavior is similar to the SLDM estimate reported in \cite{druskin1989two, druskin1994spectral}. However, as shown in \cite{saputra2026udar} {and in the section ``Comparison with other solvers''}, the proportionality coefficient for the block-Gauss quadrature is significantly improved, and unlike SLDM, its convergence is steady and monotonic, and it comes with a tight, computable stopping criterion. The combined Gauss/Gauss–Radau averaging also yields a substantial reduction in the approximation error. Moreover, potential extensions based on the Krein–Nudelman framework from \cite{druskin2026} and Ny\-ström preconditioning from \cite{chen2025Nystrom} could further reduce the dependence on the discretization parameter~$h$, particularly for strongly non-uniform meshes.

\begin{figure}[htbp]
    \centering
    \includegraphics[width=0.9\columnwidth]{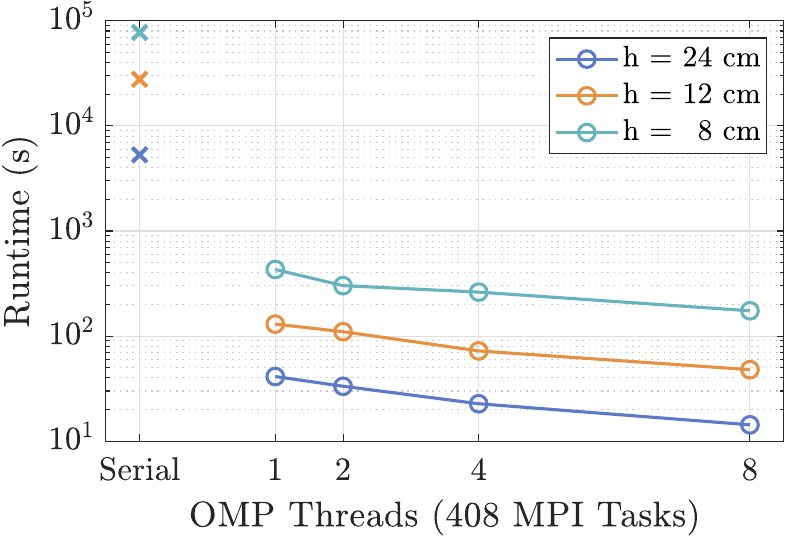}
    \caption{Dependence of (Log-)runtime for the three minimal mesh sizes on the degrees of parallelization. With the increasing number of OMP threads, the number of computing nodes was increased to have one core per thread. The $\times$-markers on the left indicate serial runtime for the three color-coded mesh sizes.}
    \label{fig:RuntimeRanks}
\end{figure}

\subsection{Hybrid OpenMP/MPI Parallelization}\label{sec:HPC}

To achieve real-time simulation of 3D UDAR measurements, a hybrid OpenMP/MPI parallelization was implemented. The MPI layer is used to launch and distribute multiple simulation tasks corresponding to different logging point–receiver configurations across the available computing resources, with multiple MPI ranks assigned per node. Within each rank, OpenMP (OMP) provides shared-memory parallelism for a single logging point across the available CPU cores. To achieve peak performance, the number of MPI tasks (ranks) should ideally match the total number of independent configurations simulated, i.e., the number of logging points multiplied by the number of receiver spacings.  The second parallel region operates at the block-quadrature-solver level using OpenMP, which optimizes numerous linear algebra operations (LAPACK/BLAS).

Table~\ref{tab:Runtime} summarizes the total runtimes obtained for 3 different mesh sizes and varying degrees of parallelism. Parallel simulations were executed on the TACC Stampede~3 supercomputer using 408 MPI tasks (one task per logging-point-receiver configuration) with 1, 2, 4, and 8 OpenMP threads per task, using 4, 8, 15, and 30 compute nodes, respectively. Serial runs were performed on a laptop for reference using 12 OpenMP threads. As expected, a consistent reduction in wall-clock time is observed with increasing thread count, as shown in Figure~\ref{fig:RuntimeRanks}. For all mesh resolutions, the runtime decreases by approximately a factor of 2.5 when moving from one to eight OpenMP threads, reflecting the combined effects of memory bandwidth limitations and coordination between OpenMP threads. When the number of cores is limited, increasing the number of MPI tasks should be prioritized over increasing the number of OpenMP threads. Logging points in resistive sections converge more slowly, so the MPI run is limited by the slowest logging point.

\begin{table}[htbp]
\caption{Runtime comparison showing strong scaling for different minimum mesh sizes and parallel configurations for the full UDAR measurements containing 408 logging points. Parallel runs were performed on dual-socket Intel\textsuperscript{\textregistered} Xeon\textsuperscript{\textregistered} CPU Max 9480 (Sapphire Rapids) nodes on TACC Stampede~3, each with 112 cores per node. Runs with 4, 8, 15, and 30 nodes were configured with 1, 2, 4, and 8 OpenMP threads per MPI task, respectively.}
\label{tab:Runtime}
\centering
\resizebox{\columnwidth}{!}{%
\begin{tabular}{|c|cccc|c|}
\hline
\textbf{min.} & \multicolumn{4}{c|}{\textbf{Parallel$^\ddagger$ Runtime (s)}} & {\textbf{Serial$^\dagger$}} \\
\cline{2-5}
 \textbf{Mesh}& \multicolumn{4}{c|}{408 MPI Tasks, varying OMP threads} &\textbf{Run}\\
\cline{2-5}
 \textbf{Size}& 1 OMP & 2 OMP & 4 OMP & 8 OMP &  \textbf{Time(s)}\\
\hline
$h = 24$ cm & 41.2 & 33.3 & 22.7 & 14.3 & \phantom{1}5,304 \\
$h = 12$ cm & 130  & 110  & 72.2 & 47.9 &   27,744 \\
$h = \phantom{1}8$ cm & 430  & 302  & 262  & 174  & 77,520 \\
\hline
\multicolumn{6}{l}{\footnotesize $\ddagger$ Supercomputer: TACC Stampede~3 - (Sapphire Rapids).} \\
\multicolumn{6}{l}{\footnotesize $\dagger$ Personal laptop: Intel\textsuperscript{\textregistered}~Core\textsuperscript{\texttrademark}~Ultra~7~165U~(1.70~GHz) 12~OMP~threads}
\end{tabular}}
\end{table}

\section{Comparison with other solvers}
We compare the proposed method with SimPEG \citep{cockett2015simpeg,heagy2017framework} in two tests. The first test concerns speed and memory. We implemented a Python prototype of the block quadrature solver that plugs into SimPEG and compared it with SimPEG's own direct solve (MKL PARDISO) on the same finite-volume discretization, so the two runs differ only in the solver. Because the prototype is interpreted, its absolute timings are conservative. For reference, we also report the matrix-free Fortran implementation used in the previous sections. The second test concerns accuracy. We simulate a full log with the Fortran code (Lebedev grid with material averaging and the block quadrature solver) and with SimPEG, and compare the two logs. We do not compare computation times in this second test, because the discretizations differ: SimPEG has no material-averaging scheme, so it is run on a mesh fine enough to converge.

\begin{figure}[t]
\centering
  \includegraphics[width=\columnwidth]{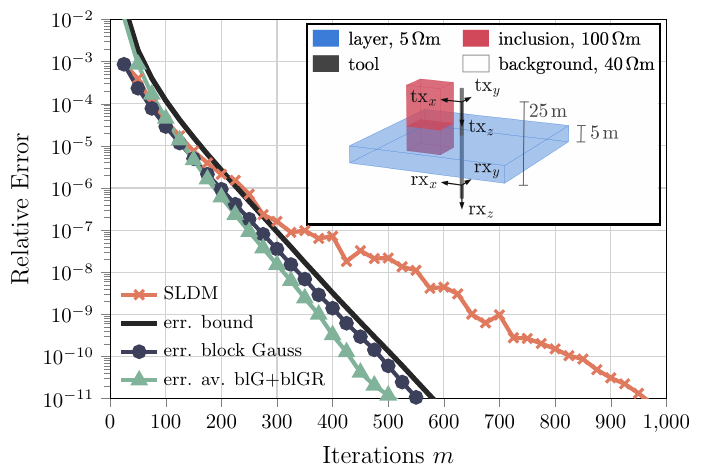}
\caption{Convergence of block Gauss, block Gauss--Radau, their average, the computable error bound, and standard Lanczos (SLDM), against a direct-solve reference, for a mesh with 770{,}894 edges. Simulated configuration shown in the top right.}
\label{fig:convergence3}
\end{figure}

The isotropic model is shown in the inset of Figure~\ref{fig:convergence3}. It consists of a 5\,m thick, 5\,$\Omega$\,m layer in a 40\,$\Omega$\,m background, and a 100\,$\Omega$\,m block measuring 12\,m\,$\times$\,10\,m\,$\times$\,20\,m in $x$, $y$ and $z$. We place the origin on the well axis at the center of the layer, with $z$ pointing down. The block then occupies $9\le x\le 21$\,m, $3\le y\le 13$\,m and $-15\le z\le 5$\,m. Its nearest vertical edge is 9.5\,m from the well. The block cuts through the layer, reaching 12.5\,m above its top and 2.5\,m below its base, and because it is offset in all three directions the response has no residual symmetry. The tool carries a triaxial magnetic-dipole transmitter 12.5\,m above the tool center and a triaxial receiver 12.5\,m below it. The vertical well is logged at 1\,m intervals, with the tool center moving from $z=-40$\,m to $z=30$\,m (71 positions), so the log covers more of the side to which the block extends. The timing test uses 6, 12, 24 and 48\,kHz, with the mesh padded to at least three skin depths at the lowest frequency in the most resistive material.

Figure~\ref{fig:convergence3} shows the relative error against PARDISO on a mesh with 770{,}894 edges. The computable error bound is tight throughout, and averaging block Gauss (blG) and block Gauss--Radau (blGR) improves on either rule alone. The curves are for the full $\calF$, which is dominated by the large self-coupling entries and therefore converges faster than the weaker transmitter--receiver coupling. The iteration counts that matter in practice are measured on the coupling block and listed in Table~\ref{tab:series3-results}. For comparison, we also ran SLDM independently for each of the six columns and averaged the errors. Even with this averaging, which smooths the curve more than the usual three runs would, the SLDM error is not monotonic and needs about twice as many iterations as the block-quadrature solver. SLDM also has no tight computable bound, so it is hard to know when to stop the iteration.

Table~\ref{tab:series3-results} compares PARDISO with three implementations of the block Gauss iteration on five mesh refinements and one deliberately oversized mesh. The first implementation is plain NumPy, using SciPy's single-threaded sparse matrix--vector product. The second is identical except that this product goes through MKL's threaded sparse BLAS, which separates the effect of multithreading from that of the algorithm. The third runs the entire iteration on the GPU with CuPy. For PARDISO, the symbolic factorization (ordering) is computed only once, since the sparsity pattern is the same for all four frequencies, and only the numeric factorization is repeated per frequency. PARDISO memory is its self-reported numeric-factorization peak for a single factorization. Block Gauss memory counts four block vectors of size $N\times p$, i.e., the three Lanczos vectors plus one work vector of the Python prototype; as for PARDISO, the system matrix is not included. Even the NumPy version is 5.6 to 14.6 times faster than PARDISO, the MKL version 8.5 to 23.7 times and the GPU version 22 to 77 times, and the speedup grows with mesh size. Block Gauss also needs 75 to 258 times less memory. On the largest mesh (4.6 million edges), PARDISO ran out of memory on a 256\,GiB machine, while block Gauss completed using 847\,MiB. On the mesh with 2.0 million edges, the matrix-free Fortran code takes about 78\,s, compared with 3625\,s for PARDISO.

\begin{figure}[t]
\centering
  \includegraphics[width=\columnwidth]{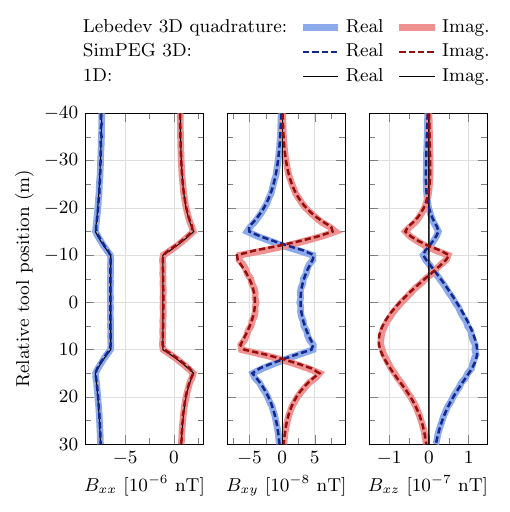}
\caption{Logs of the $B_{xx}$, $B_{xy}$ and $B_{xz}$ couplings versus tool position (tool center) relative to the center of the 5~$\Omega$\,m layer. The proposed Lebedev 3D quadrature solver is compared with SimPEG and with the 1D (semi-)analytical solution; the 1D model does not include the 3D resistive block.}
\label{fig:bfield_toprow}
\end{figure}

Figure~\ref{fig:bfield_toprow} compares the logs at 6\,kHz. For brevity, we show only the couplings of the $x$ transmitter to the three receiver components, $B_{xx}$, $B_{xy}$ and $B_{xz}$. SimPEG was run on a tool-centered mesh with 2\,m horizontal and 0.5\,m vertical core cells, giving 1.84 million edges per position. The Fortran code and SimPEG are in excellent agreement along the whole log. We also show the 1D (semi-)analytical solution for the model without the block. For a vertical well in a horizontally layered medium, the cross-couplings $B_{xy}$ and $B_{xz}$ vanish by symmetry, so their non-zero values in the 3D logs come entirely from the block, which shows that the response is genuinely three-dimensional.

\begin{table*}[t]
\caption{Wall-clock time for the block Gauss solver versus SimPEG's direct solve via MKL PARDISO, across mesh refinements of the induction-logging test case. Plain NumPy compared to the same code with only its sparse matmul routed through MKL, and a separate full CUDA/CuPy implementation. All configurations ran with 16 threads available to MKL/PARDISO. Peak memory for block Gauss does not depend on backend (it depends only on $N$ and $p$). Speedup columns are PARDISO time (resp.\ memory) divided by the corresponding block Gauss figure.}
\label{tab:series3-results}
\centering
\resizebox{\textwidth}{!}{%
\begin{tabular}{r rrrr rr r rrr r r}
\toprule
 & \multicolumn{4}{c}{Block Gauss} & \multicolumn{2}{c}{SimPEG / PARDISO\textsuperscript{a}} & Fortran & \multicolumn{3}{c}{Speedup vs.\ PARDISO} & mem.\ & Iter.\textsuperscript{d} \\
\cmidrule(lr){2-5} \cmidrule(lr){6-7} \cmidrule(lr){9-11}
$N$ (edges) & NumPy [s] & MKL\textsuperscript{b} [s] & CuPy [s] & mem [{\bf MiB}] & $t$ [s] & mem [{\bf GiB}] & $t$ [s]\textsuperscript{c} & NumPy & MKL & CuPy & adv. & $m_{\scriptstyle 10^{-6}}$\\
\midrule
82{,}969 & 1.75 & 1.18 & 0.45 & 15.2 & 9.99 & 1.11 & --- & 5.7$\times$ & 8.5$\times$ & 22.2$\times$ & 75$\times$ & 125 \\
287{,}584 & 14.45 & 7.88 & 2.31 & 52.7 & 81.17 & 6.12 & --- & 5.6$\times$ & 10.3$\times$ & 35.1$\times$ & 119$\times$ & 250 \\
770{,}894 & 75.33 & 38.85 & 12.31 & 141.2 & 491.2 & 24.16 & --- & 6.5$\times$ & 12.6$\times$ & 39.9$\times$ & 175$\times$ & 525 \\
1{,}254{,}144 & 138.0 & 78.04 & 23.76 & 229.6 & 1313.4 & 47.37 & --- & 9.5$\times$ & 16.8$\times$ & 55.3$\times$ & 211$\times$ & 650 \\
2{,}023{,}623 & 248.8 & 153.2 & 46.99 & 370.5 & 3625.1 & 93.38 & 77.8 & 14.6$\times$ & 23.7$\times$ & 77.2$\times$ & 258$\times$ & 800 \\
4{,}622{,}940 & 806.7 & 564.3 & 149.0 & 846.5 & $\times$ & $>256$ & --- & $\times$ & $\times$ & $\times$ & $\times$ & 1200 \\
\bottomrule
\end{tabular}}

\vspace{3pt}
{\footnotesize
\begin{minipage}[t]{0.6\textwidth}
\raggedright
\textsuperscript{a}\,PARDISO's self-reported numeric-factorization peak (\texttt{iparm}).\\
\textsuperscript{b}\,Identical code; only the sparse matvec is routed through MKL instead of SciPy.\\
\textsuperscript{c}\,Fortran implementation used in all other sections, same iteration count.
\textsuperscript{d}\,Iterations for the error bound to reach coupling tolerance $10^{-6}$ at 6\,kHz, the slowest-converging frequency.\\
\end{minipage}\hfill
\begin{minipage}[t]{0.38\textwidth}
\raggedright
$\times$\,PARDISO ran out of memory.\\
CPU: Intel\textsuperscript{\textregistered} Xeon\textsuperscript{\textregistered} w5-3435X (16 physical cores), 256\,GiB RAM, 16 threads.\\
GPU: NVIDIA RTX 4000 Ada Generation (20\,GiB).
\end{minipage}}
\end{table*}

\section{Conclusions}
We developed, implemented, and verified a new goal-ori\-ented ROM framework specifically targeted at far-zone EM measurements, such as borehole UDAR arrays.
The fundamental technical innovation of this approach is treating the problem as block-quadrature targeted to trans\-mitter-\-receiver coupling simulations rather than EM-field approximation on the entire FD grid.
Our new simulation framework builds on the authors' recent contributions, including fully automated heterogeneous media averaging and an accelerated, robust block-quadrature solver optimized for modern HPC architectures.
The solver provides tolerance control through quadrature error bounds and enables cost-free error reduction through quadrature averaging.

Moreover, the solver supports the computation of adjoint-based Jacobians for gradient-based inversion workflows with only a negligible increase in computational cost.
Large-scale numerical simulations of a synthetic set of UDAR measurements verified the efficiency, scalability, and competitiveness of the developed approach and its MPI-based HPC implementation. A comparison with SimPEG confirmed the accuracy of the method and showed substantial speed and memory advantages of the block-quadrature solver over a direct solver.

When implemented on massively parallel computers, the algorithm can accurately simulate UDAR measurements acquired along a large-scale 3D anisotropic formation comprising an array of five transmitter–receiver separations, each including nine triaxial couplings and operating at four discrete frequencies. Such a configuration, corresponding to a single logging point along the well trajectory, can be simulated in tens of CPU seconds. For one logging point, the block-quadrature solver converges in half the iterations of SLDM at points where SLDM converges well and computes all couplings at once. In spatially complex geological settings, however, SLDM may exhibit erratic behavior and converge slowly, whereas the proposed approach remains robust and converges monotonically.

This forward modeling approach has already enabled fast 2D and 3D (hierarchical) inversion of UDAR measurements and allows for detailed pre-drilling studies. Future work will focus on improving the averaging scheme to better capture corner effects, thereby enhancing accuracy in complex geological settings and near faults. We also plan to accelerate the quadrature solver using a GPU implementation, randomized preconditioners, and Krein--Nudelman approaches to block quadratures. Preliminary experiments suggest speedups of at least one order of magnitude, bringing real-time and ahead-of-the-bit 3D inversion of UDAR-LWD measurements with stochastic uncertainty quantification within reach.

Beyond UDAR, the proposed targeted quadrature framework should extend to controlled-source electromagnetics (CSEM) and other geophysical problems with limited observation sets governed by parabolic PDEs, as discussed in the Introduction.

\section*{ACKNOWLEDGMENTS}

This work was supported by the University of Texas at Austin’s Deep Imaging Sub-Consortium (of the Research Consortium on Formation Evaluation),
currently sponsored by AkerBP, Baker-Hughes, bp, ENI, Equinor ASA, Halliburton, and Petrobras. The authors also acknowledge the Texas Advanced Computing Center at the University of Texas at Austin for providing HPC resources on Stampede 3.

Carlos Torres-Verdín is thankful for the financial support provided by (a) the Brian James Jennings Endowed Memorial Chair in Petroleum and Geosystems Engineering, and (b) the 2025 Moncrief Grand Challenge Award by the Oden Institute, both from the University of Texas at Austin.

Vladimir  Druskin was partially supported by AFOSR grants FA 9550-20-1-0079, FA9550-23-1-0220,  and NSF grant  DMS-2110773.

\section*{Data availability}

The geological models and data associated with this research are confidential and cannot be released. The synthetic model used in the comparison with SimPEG is fully specified in the manuscript. The production source code is proprietary and cannot be published; a Python prototype of the quadrature solver for SimPEG is available from the corresponding author on request for academic use, and the authors are glad to assist researchers implementing the method from the formulations provided.

\bibliographystyle{seg}
\bibliography{bib}

\end{document}